\documentclass[aip,reprint,twocolumn]{revtex4-1}

\usepackage{graphicx}

\begin{document}

\preprint{\today}

\title{The study of the microwave heating of bulk metals without microwave susceptors}


\author{Maxim Ignatenko}
 \email{mignat@isc.chubu.ac.jp}
 \affiliation{Chubu University,1200, Matsumoto-cho, Kasugai-shi 487-8501, Japan.}
\author{Motohiko Tanaka}
 \affiliation{Chubu University,1200, Matsumoto-cho, Kasugai-shi 487-8501, Japan.}


\date{\today}

\begin{abstract}
{This paper discusses the physical aspects of the microwave heating of bulk metals in multimode cavity. The concept of power balance and  thermodynamics are utilized to explain the preheating step required for successful microwave heating.}
\end{abstract}

\maketitle 
%
%

The utilization of microwaves for processing of metal powders and metal parts is an unexpected and rapidly developing method. Compared to conventional methods, tt improves quality of products and reduces processing time\cite{Roy1999}. The most striking experimental result is the possibility to sinter metal parts directly by microwaves without the assistance of microwave susceptors\cite{Anklekar2001,Anklekar2005,Agrawal2010}.
\par
It is well known that good electrical conductors reflect most of incident microwave power. For example, semi-infinite copper slab irradiated at frequency $f=2.45$~GHz reflects about $99.98$\% of incident power. The rest of power is dissipating owing to Joule loss. The high reflectivity is caused by significant macroscopic current induced by the time varying magnetic field component. In metal powders the current is suppressed, and this promotes the microwave heating\cite{Cheng2001, Ignatenko2010, Ignatenko2011}. In turn, bulk metals can be heated by microwaves with assistance of microwave susceptors, which convert microwave energy into thermal energy, and heat surroundings in conventional way. Owing to high reflectivity, the metal parts by microwaves without susceptors was unbelievable up to recent experiments. 
\par
\par


First theoretical explanations of microwave experiments with metal powders and parts utilized surface impedance concept and assumed the use of either microwave susceptors or perfect thermal insulation\cite{Mishra2006,Luo2004}. They had demonstrated the principal possibility of the microwave heating of metal parts but didn't provide insight into the conditions of the above experiments. This paper presents another approach. Instead of calculations of Maxwell's equations coupled with the heat equation, it uses the concept of energy balance to scan the parameters affecting the heating. This scan points to the conditions favorable for successful experiments and outlines the region of interest for further investigations.
\par
The concept of energy balance is originated from the heat equation:
\begin{equation}
  \rho C_{\rho} \frac{dT}{dt} = \nabla\left(\kappa\nabla T\right) + P_{+}(\textbf{r},T) - P_{-}(T_{\textrm{s}}).
\label{eq2.1}
\end{equation}
Here, $T$ is the sample temperature; $\rho$, $C_{\rho}$ and $\kappa$ are the mass density, specific heat capacity and thermal conductivity of the sample, respectively; $P_{+}$ is the volume density of dissipating microwave energy per unit time at space point $\textbf{r}$ and $P_{-}$ determines the heat loss through the sample surface due to radiation and convection
\begin{equation}
  P_{-}(T_{s}) = A e_{\textrm{rad}}\sigma_{\textrm{SB}}\left(T_{\textrm{s}}^4 - T_{\textrm{surr}}^4\right)
                      + A h \left(T_{\textrm{s}} - T_{\textrm{surr}}\right),
\label{eq2.2}
\end{equation}
with $A$ is the sample area to volume ratio, $e_{\textrm{rad}}$ is the emissivity, $\sigma_{\textrm{SB}} = 5.67\times10^{-8}$~J/(s$\cdot$m$^2\cdot$K$^4$) is the Stefan-Boltzmann constant, and $h$ is the convection heat transfer coefficient; the surrounding temperature $T_{\textrm{surr}}$ and the surface temperature $T_{\textrm{s}}$ are in the Kelvin scale. 
\par
The diffusion term tends to equalize the temperature profile inside of the sample, and it can be neglected when the temperature is almost uniform. Non-uniform temperature complicates heat transport analysis and makes questionable the basis of the approach used in this paper. Fortunately, in small metal parts as well as low-loss dielectrics, variation of the temperature profile is significantly reduced. 
\par
During microwave heating, the main source of non-uniform temperature profile is the non-uniform source of heat. The variation of temperature $\Delta T$ due to the variation of dissipating power $\Delta P_{+}$ is estimated as \cite{Bykov2001}
\begin{equation}
  \Delta T \sim T\frac{\Delta P_{+}} {P_{+}} \frac{\tau_s}{\tau_T},
\label{eq2.3}
\end{equation}
where $\tau_T \sim T/(\partial T/\partial t)$ is the characteristic time of heating, and characteristic time scale of temperature equalization is given by
\begin{equation}
\tau_s=\frac{\Lambda^2}{\chi},
\label{eq2.4}
\end{equation}
with $\Lambda$ is the space scale of $P_{+}$ variation and $\chi=\kappa/(\rho C_{\rho})$ is the thermal diffusivity of material. 
\par
In the case of good electrical conductors $\Delta P_{+}/P_{+}\sim1$. For metal samples with size of a one centimeter, $\tau_s$ is typically within a range from 0.1s to 1s. Under these conditions, the heating rate slower than 1$^0$C/s causes temperature non-uniformity less than 1$^0$C, and therefore the diffusion term in Eq.~(\ref{eq2.1}) can be ignored. In the case of ellipsoidal low-loss dielectrics, $\Delta P_{+}\sim0$ and therefore temperature non-uniformity is zero.
\par
The negligibly small diffusion term allows to formulate the statement of the concept of energy balance: the saturated temperature is determined by balance between heat source $P_{+}$ and heat loss $P_{-}$, and when $P_{+} - P_{-} >0$ the sample is heated up, otherwise not. This concept significantly reduces the complexity of the problem.
\par
The paper assumes that in heat loss term $P_{-}$ radiation loss dominates and therefore $h=0$.
\par
In order to simplify the calculations of dissipating power, one need to make additional assumptions. The first assumption is that the sample size is much less than the microwave wavelength in cavity, such that the non-uniformity of the electromagnetic field is negligible and the microwave heating by a plane running wave is considered.
The next assumptions regard material properties and sample shape: permittivity and permeability are uniform, and the sample is spherical. In this case, the amount of the power dissipated in the sample is easily calculated by means of Mie theory\cite{Bohren1983}.
\par
Finally, the problem of three dimensional numerical solution of Maxwell's equations is reduced to the calculation of Mie coefficients, and the problem of heat transport is reduced to the investigation of energy balance $P_{+} - P_{-}$. 
\par


In order to calculate Mie coefficients, one need to specify optical parameters of sample. In the case of copper, the magnetic permeability is $\mu_s = \mu_0$, while the dielectric permittivity $\varepsilon_s$ is determined by the electric conductivity $\sigma$ and microwave frequency $f$ as
\begin{equation}
  \varepsilon_s = i \frac{\sigma(T)}{\omega},
\label{eq4.1}
\end{equation}
where $i=\sqrt{-1}$ and $\omega=2 \pi f$, and the temperature dependence of the electrical conductivity is given by
\begin{equation}
  \sigma(T) = \frac{\sigma_0}{1+\beta(T - T_0)},
\label{eq4.2}
\end{equation}
with $\sigma_0=5.8\times10^{7}$ 1/($\Omega\cdot$m) is the electrical conductivity at temperature $T_0=25^0$C, and $\beta = 0.0042$ is the temperature coefficient.
\par
Emissivity of the sample depends on surface conditions. The sum of emissivity and reflectivity equals to unity by Kirchoff's law. From everyday experience it is known that clear metal surface is bright while oxide layer tarnishes the surface. This points to small emissivity of clear metal compared to that of oxidized surface. For example, the emissivity of clear copper sample is taken to be 0.03, and that of oxidized  copper is 0.80 \cite{emissivity}.
\par
In calculations, sample radius is 5~mm, and the amplitudes of the electromagnetic field are typical for single-mode cavity experiments ($10^4\sim10^5$~V/m), \emph{i.e.} they are higher than typical strength of the electromagnetic field utilized in multimode cavities.
\par
\begin{figure}
	\begin{center}
	\includegraphics[width=70mm]{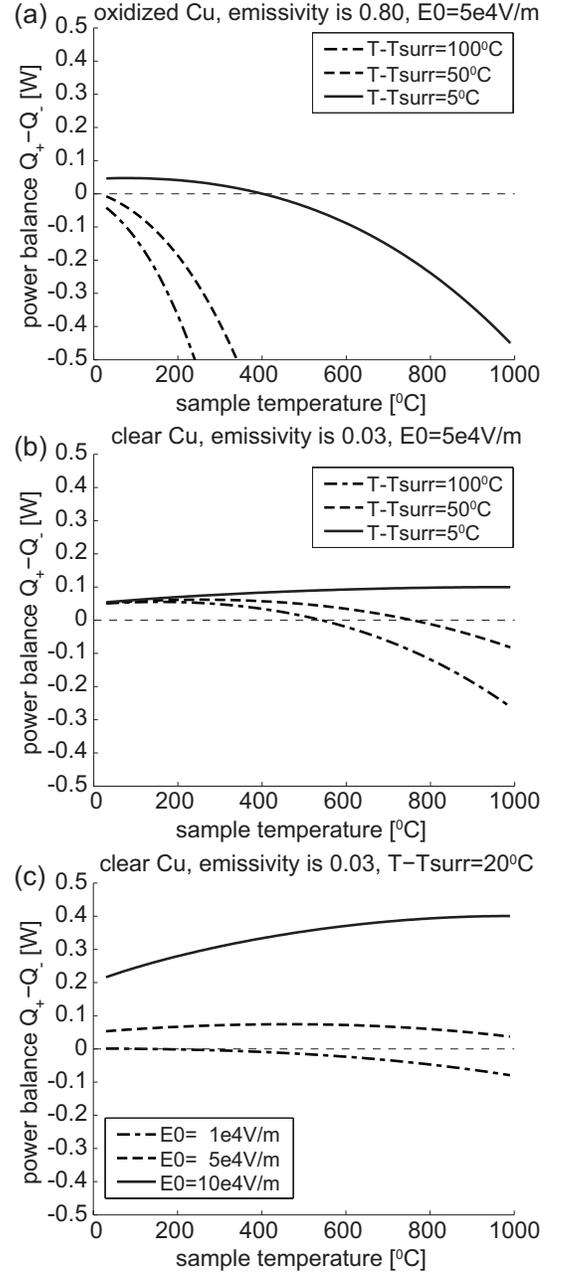}
	\end{center}
\caption{Power balance of oxidized copper sample (a) and clear samples (b,c). Here $E_0$ is the amplitude of the electric field component of incident plane wave, and $T-T_{\textrm{surr}}$ is the difference between temperature of sample and that of thermal insulation.}
\label{fig1}
\end{figure}

Power balance expressed in watts ($Q=P\cdot V_s$, where $V_s$ is the sample volume) is shown in Figure~\ref{fig1}. 
Results of calculations clearly demonstrate that oxidized copper sample is hard to heat even by strong electromagnetic field. Radiation heat loss limits the maximum temperature of the sample at 400$^0$C when $T-T_{surr}=5^0$C [Fig.~\ref{fig1}(a)]. In the case of larger loss, the sample is not heated at all.
The sample with clear surface is heated well, though the heating is still not easy and requires strong electromagnetic field [Fig.~\ref{fig1}(b,c)]. For example, if $T-T_{surr}=50^0$C, the clear sample is heated almost up to 800$^0$C when $E_0=5\times10^4$~V/m, and maximal temperature is higher than 1000$^0$C when the magnitude of the electric field increased by two times $E_0=10\times10^4$~V/m.
\par
The obtained data provide a key for understanding of experimental results. Clear sample surface significantly reduces radiation loss and correspondingly promote heating.
In experiments, the surface is cleared when the sample is preheated in reducing atmosphere or vacuum. For example, in hydrogen atmosphere the reduction of metal $M$ from oxide $M_x O_2$ is governed by
\begin{equation}
  M_x O_2(s) + 2H_2(g)= xM(s) +2H_2O(g),
\label{eq5.2}
\end{equation}
where $x$ represents the stoichiometry of the oxide; $(s)$ and $(g)$ stand for ``solid'' and ``gas'', respectively. The equilibrium ratio of hydrogen to water partial pressures $P_{{H_2}}/P_{{H_2O}}$ is determined by the change of Gibbs energy for the reaction $\Delta G$:
\begin{equation}
  {\textrm{ln}}\frac{P_{{H_2}}} {P_{{H_2O}}} = \frac{\Delta G}{RT},
\label{eq5.2}
\end{equation}
with $R=8.314$~J/(K$\cdot$mol) is the gas constant and $T$ is the absolute temperature. 
\par
\begin{figure}
	\begin{center}
	\includegraphics[width=75mm]{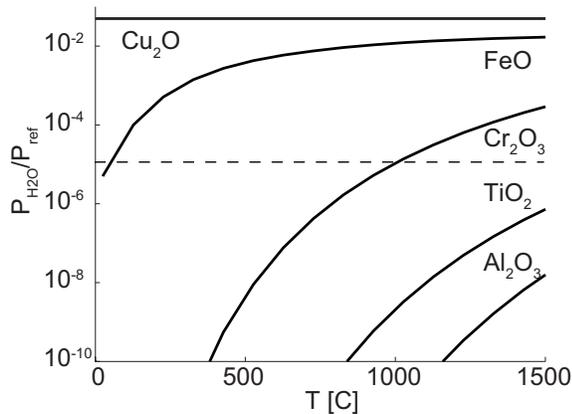}
	\end{center}
\caption{Equilibrium partial pressure of moisture for different metals/metal oxides systems. Dashed horizontal line corresponds to moisture pressure maintained in experiments. Equilibrium partial pressure above of this line promotes the reduction reaction. Reference pressure $P_{\textrm{ref}}$ equals to $10^5$~Pa. Data is taken from JANAF thermochemical tables \cite{JANAF1985}.}
\label{fig3}
\end{figure}

Assuming that input gas atmosphere consists of 5\% $H_2$ and 95\% of neutral gas, corresponding equilibrium partial pressure of water is shown in Figure~\ref{fig3}. 
Dashed horizontal line shows the level of moisture maintained in the experiments estimated based on measured dew point $T_D=-60^0$C by means \cite{German1996}
\begin{equation}
  {\textrm{log}}_{10}(V_{{H_2O}}) = -0.24 + 0.03T_D - 1.74 \cdot 10^{-4} T_D^2 + 5.05 \cdot 10^{-7} T_D^3.
\label{eq5.3}
\end{equation}
The calculated moisture contents $V_{{H_2O}}=10^{-3}$~vol.\% corresponds to partial pressure $P_{{H_2O}}/P_{\textrm{ref}} \approx 10^{-5}$. The calculations show that preheating of copper and monoxide of iron is highly effective for the reduction of sample surface from oxide. The reduction of chrome, titanium and aluminum oxides is not so easy and requires much higher temperature and/or more stringent moisture control.
\par
In summary, the microwave heating of metal parts is possible owing to non-trivial prerequisite experimental procedure, when the surface of sample is cleared from oxides during preheating in reducing atmosphere. This procedure significantly reduces radiation heat loss, and makes possible the temperature rise, provided the strong electromagnetic field and good thermal insulation. The heat is produced by Joule loss owing to eddy current. 
\par
This paper considered the heating of bulk copper. Compared to copper, many common metals important for applications, like Ni, Fe and steels, possess lower electrical conductivity (Table~\ref{t1}). Besides, their magnetic properties are favorable for microwave heating.
\begin{table*}
\caption{Typical values of the electrical conductivity, permeability and emissivity of clear and oxidized metals at room temperature \cite{Paul2006,titan,emissivity}. Electrical conductivity of copper $\sigma_{\textrm{Cu}}$ is $5.8\times10^{7}$ 1/($\Omega\cdot$m).}
\label{t1}
\begin{center}
\begin{ruledtabular} 
\begin{tabular}{lccccccc}
                                               & Cu & Al & Ni & Fe & Steel & Stainless steel & Ti \\
\hline
$\sigma/\sigma_{\textrm{Cu}}$ & 1 & 0.61 & 0.20 & 0.17 & 0.10  & 0.02 & 0.04 \\
\hline
$e_{\textrm{rad}}$ clear           & 0.03 & $\sim$0.05 & 0.07      & 0.1 - 0.4 & 0.07 & 0.08 - 0.8 & 0.1 - 0.2 \\
$e_{\textrm{rad}}$ oxidized        & 0.8   & 0.4 - 0.6    & 0.6 - 0.9 & 0.3 - 0.9 & 0.8  &  -            & 0.5 - 0.6 \\
\end{tabular}
\end{ruledtabular} 
\end{center}
\end{table*}
They dissipate more microwave power, and thus they are good candidates for microwave heating experiments. In addition, Fe and Ni are reduced from oxides easily in hydrogen atmosphere (diagram for Ni is not shown).

This work was performed under the support of a Grant-in-Aid for Scientific Research on Priority Area No.18070005 (FY2006-2010) from the Japan Ministry of Education, Culture, Sports, Science and Technology.


\begin{thebibliography}{99}
\bibitem{Roy1999}
			R.~Roy, D.~Agrawal, J.~Cheng, and S.~Gedevanishvili:
			Nature \textbf{399} (1999) 668.
\bibitem{Cheng2001} 
			J.~Cheng, R.~Roy, and D.~Agrawal: 
			J. Mater. Sci. Lett. \textbf{20} (2001) 1561.
\bibitem{Ignatenko2010}
			M.~Ignatenko and M.~Tanaka:
			Physica B \textbf{405} (2010) 352.
\bibitem{Ignatenko2011}
			M.~Ignatenko and M.~Tanaka:
			submitted to JJAP.
\bibitem{Anklekar2001}
			R.M.~Anklekar, D.K.~Agrawal and R.~Roy:
			Powder Metallurgy \textbf{44} (2001) 355.
\bibitem{Anklekar2005}
			R.M.~Anklekar, K.~Bauer, D.K.~Agrawal and R.~Roy:
			Powder Metallurgy \textbf{48} (2005) 39.
\bibitem{Agrawal2010}
			D.~Agrawal, private communication
\bibitem{Mishra2006}
			P.~Mishra, G.~Sethi, and A.~Upadhyaya:
			Metall. Mater. Trans. B \textbf{37B} (2006) 839.
\bibitem{Luo2004}
			J.~Luo, C.~Huyar, L.~Feher, G.~Link, M.~Thumm, and P.~Pozzo:
			Appl. Phys. Lett. \textbf{84} (2004) 5076.
\bibitem{Bykov2001} 
			Yu.~V.~Bykov, K.~I.~Rybakov, and V.~E.~Semenov: 
			J. Phys. D: Appl. Phys. \textbf{34} (2001) R55.
\bibitem{Bohren1983}
                    C.~F.~Bohren and D.~R.~Huffman:
                    \textit{Absorption and Scattering of Light by Small Particles} (Wiley, New York, 1983).
\bibitem{emissivity}
			http://www.engineeringtoolbox.com/emissivity-coefficients-d\_447.html
\bibitem{JANAF1985} 
			M. W. Chase, Jr., C. A. Davies, J. R. Downey, Jr., D. J. Frurip, R. A. McDonald, and A. N. Syverud:
			\textit{JANAF Thermochemical Tables} (National Institute of Standards and Technology, Washington, 1985) 3rd ed.
 \bibitem{German1996} 
			R.M.~German:
			\textit{Sintering Theory and Practice} (Wiley, New York,1996).
\bibitem{Paul2006} 
			C.R.~Paul:
			\textit{Introduction to Electromagnetic Compatibility} (Wiley, New Jersey, 2006) 2nd ed.
\bibitem{titan}
			http://en.wikipedia.org/wiki/Titanium


%
%
%
\end{thebibliography}
\end{document}